\documentstyle[aps,epsfig,multicol,fancyheadings]{revtex} 
\begin{document}
\draft
\title{Ground state non-universality in the random-field Ising model}
\author{P.M. Duxbury and J.H. Meinke \footnote{ \noindent 
{\bf email:} meinke@pa.msu.edu}}

\address{ Dept. of Physics and Astronomy and 
Center for Fundamental Materials Research,\\
Michigan State University, East Lansing, MI 48824, USA.
}
\maketitle

\begin{abstract}
Two attractive and often used ideas, namely 
universality and the concept of a zero temperature
fixed point, are violated in the infinite-range
random-field Ising model.   In the ground state we 
show that the exponents can depend {\it continuously} 
on the disorder and so are non-universal.
However, we also show that {\it at finite temperature} the
thermal order-parameter exponent $1/2$ is restored so that temperature is
a relevant variable. The broader implications of these
results are discussed.
\end{abstract}

\pacs{75.10.Nr, 61.43.Bn, 64.60.Ak}

\begin{multicols}{2}[]
\narrowtext

Co-operative behavior in disordered systems 
can usually be concisely characterised using scaling theories.
  These scaling theories contain scaling exponents and 
  it is particularly important and satisfying if these 
  exponents are independent of the fine details of the model, 
  that is they are in some sense ``universal''.  
  If this holds it allows the theorist to study 
  the simplest or most convenient model in a class 
  in order to find the scaling exponents and, more 
  importantly, that experiments should show the same 
  exponents as the theory even though they may look 
  very different on short length scales.  Universality
   has been spectacularly successful in the study of 
   phase transitions as a function of temperature, 
   culminating in the development of the renormalisation 
   group\cite{wilson}.  There the critical exponents, usually, only 
   depend on the symmetry of the order parameter and 
   the spatial dimension.
    Due to the fact that scaling theories also work in 
    many disordered systems it is natural to try to extend 
    the ideas which work so well for thermal 
    phase transitions to the disorder case.

  Universality in disordered systems assumes that the
  critical exponents should {\it not depend of the type of disorder},
  provided the disorder distribution is short-range correlated 
  and provided it is not too broad.  Universality with respect to 
  disorder has been confirmed in  some systems, with 
  notable examples being percolation\cite{stauffer} and the problem 
  of a directed polymer in a random medium\cite{halpin}.
  However universality in disordered systems  is unproven in
  general and indeed it has been questioned in the spin glass problem 
  where there appears to be qualitative difference 
  between the behavior in the presence of Gaussian
   as compared to bimodal disorder\cite{young}.   More recently
   universality has even been questioned in the random field
   Ising model\cite{swift,angles}, which is one of the simplest models of a 
   disordered material.  Moreover the experimental tests of the
   random field Ising exponents rely on universality\cite{fishman,cardy}
   as the experiments are carried out on {\it diluted antiferromagnets
   in a field}\cite{belanger} which 
   are expected to lie in the same universality class 
   (these experiments are also plagued by kinetic effects due to the
   large barriers which exist in random magnets).   
    We show that universality fails in the 
    ground state of the infinite-range
     random-field Ising model, as the {\it exponents may vary continuously} 
     with the type of disorder.  
        
    However, we also show that universality is restored at any
    finite temperature in the sense that at finite temperature the
    order parameter exponent is always $1/2$, when the transition is continuous. 
      This implies that another important
    concept in disordered systems, the
    concept of a {\it zero temperature fixed point}, is
    violated in this model. 
   The origin of the concept of a {\it zero temperature fixed point}
   is that disorder usually provides a
   stronger perturbation than thermal fluctuations 
   (which in turn are usually a stronger perturbation
   than quantum fluctuations).  Thus a study of the 
   {\it ground states of disordered systems} can lead to 
    scaling theories which are qualitatively 
    correct at finite temperatures.  This is particularly 
    attractive since there now exist methods for finding 
    the exact ground states of many quenched random systems\cite{alavadomb}. 
     In the vernacular of random systems it is 
      often stated that there exists a  
     zero-temperature fixed point which controls
      the behavior at finite temperature.   
       In particular, in both the random field Ising model
        \cite{bray,villain,fisher,ogielski}
       and in spin glasses\cite{young}, scaling theories
       are frequently based on the  
       assumption of a zero-temperature fixed point. 
        However, we show that the mean-field theory 
        of the random field Ising model  
   {\it is not, in general, controlled by a zero-temperature fixed point}.

We first demonstrate that the critical exponents can take on
a range of values in the ground state of the random-field Ising model. 
 The Hamiltonian for this model is,
\begin{equation}
{\cal H} = -J_0 \sum_{ij} S_i S_j - \sum_i h_i S_i = N E_{ex} + N E_f
\end{equation}
where the first sum is over all spin pairs and 
$J_0=J/N$ where $N$ is the number of sites in the lattice
to ensure an extensive energy.  
 When the distribution of random fields is narrow, 
 the exchange term dominates and the system is a ferromagnet
(in dimensions greater than or equal to three), 
while when the random field distribution is broad 
the random field dominates and the system becomes a 
paramagnet.  We take distributions of random fields
 which have mean zero and width $\delta h$, and 
 consider the key ratio $H=\delta h/J$ which measures 
 the strength of the random field in comparison to the exchange.

Consider a spin subspace in which the 
magnetisation, $m$, is fixed, ie. 
$m = (n_+ - n_-)/N$, where $n_+$ is 
the number of up spins in the configuration 
and $n_-$ is the number of down spins.  It is 
easy to find the lowest energy state for 
fixed $m$.  The exchange energy is given by,
\begin{equation}
E_{ex}(m) = {-J \over 2 N^2} (n_+^2 + n_-^2 - 2 n_+n_-)
 = {-J m^2\over 2} 
\end{equation}
Due to the fact that the exchange is of infinite range, 
all configurations at fixed $m$ have the same energy
 and so are combinatorially degenerate.  The field 
 term splits this degeneracy by choosing the configuration 
 which has the smallest field energy.  This is achieved by 
 satisfying the largest random fields and leaving the 
 smallest possible fields unsatisfied.  If the distribution
  of random fields is $P(h)$ (which we assume to be symmetric
   about the origin), then in the large lattice limit, we have,
\begin{equation}
E_f = - 2\int_0^{\infty}dh h P(h) + 2\int_0^{h_c(m)}dh hP(h)
\end{equation}
where the first term is the ideal field energy 
in which every spin is oriented
in the direction of its local field, and the 
second term is the energy cost due to the fraction
 of fields which are unsatisfied.  The fraction of 
 fields which are unsatisfied is determined by the magnetisation,
\begin{equation}
m = 2\int_0^{h_c(m)} P(h) dh.
\end{equation}
The ground state is found by determining the value of $m$ which minimizes 
the energy (1-3), given the constraint (4). Carrying out the variation 
yields,
\begin{equation}
{\partial (E_{ex} + E_{f})\over \partial m} = - J m 
 + 2 h_c(m) P(h_c(m)) {\partial h_c(m) \over \partial m}.
\end{equation}
By taking a derivative of Eq.\ (4) with respect to 
$m$ (using the chain rule) we find, $1 = 2 P(h_c(m)) \partial h_c(m)/\partial m$
Using this to remove $\partial h_c(m)/\partial m$ from the RHS 
of Eq.\ (5) and  setting Eq.\ (5) to zero, we find that 
the cutoff field is related to the magnetisation via $h_c(m) = Jm$. 
 Subsitution of this into Eq.\ (4) yields the {\it ground-state mean-field equation}
\begin{equation}
m = 2\int_0^{Jm} P(h) dh.
\end{equation}
This equation gives the magnetisation values at which the energy is extremal.
Note that $m=0$ (the paramagnet) is always an extremum, as expected.  Since we
are treating the case of symmetric random fields, we can restrict attention
to the case where $0\le m \le 1$. To 
determine whether an extrema is a maximum or a minimum, we need to evaluate
the curvature near the extremum,
\begin{equation}
{\partial^2 (E_{ex} + E_{f})\over \partial m^2}|_{m_s} =
  -J  + { 1\over 2 P(h_c(m_s))} 
 \end{equation}
 Finally, in order to determine the ground state, we need 
 to compare the free energies of the solutions to the
  mean-field equation (4) with the energy of the magnetised state ie. $m=1$.
 
An elegant result due to Aharony\cite{aharony} states that the nature of the
finite temperature phase transition in the infinite range random field Ising model
depends on the curvature of the disorder distribution at the origin.
Bimodal distributions lead to a first order jump in the order parameter
 at low temperatures(and hence a tricritical point at finite temperature),
  while unimodal distributions exhibit continuous transitions with
  exponent $\beta = 1/2$, as originally found by Schneider
   and Pytte\cite{schneider} for the
  case of Gaussian disorder. 
   However, we now show that the 
   exponent $\beta = 1/2$ is {\it not universal} in the
   ground state.

We show that $\beta$ may change continuously with the disorder, 
by considering the distribution of random fields given by,
\begin{equation}
P(h) = {y+1 \over 2 y H} ( 1 - ({|h|\over H})^y)\ \ \ -H\le h \le H,
\end{equation}
with $y\ge 0$. In the limit $y\rightarrow\infty$ $P(h) 
\rightarrow uniform$ so that a first order behavior 
is expected, while if $y\rightarrow 2$ it looks like
 a Gaussian near the origin so we expect a continuous 
 transition with $\beta = 1/2$. In the following discussion, 
 we take $J=1$, so that $H$ has been normalised by $J$. 
  The distribution (8) is the first two terms in the expansion
  of the stretched exponential, ${\rm exp}(-(|h|/H)^y)$ which has the
  same critical behavior as (8).  
However, Eq.\ (6) can be solved exactly 
for the case (8) to yield (in addition to $m=0$),
\begin{equation}
m_s = H (y(H_c -H))^{1/y}\ \ \ {\rm for } \ \ 1\le H \le H_c,
\end{equation}
where the critical field is given by, $H_c = (y+1)/y$. 
The lower bound on $H$ is due to the cutoff in Eq.\ (8).  
For $H<1$, the exchange always wins and the magnetisation is $m=1$. 
 From (7), the second derivative is, 
\begin{equation}
{\partial^2 (E_{ex} + E_{f})\over \partial m^2}|_{m=0} =
 -1 + { y H \over y+1} 
 \end{equation}
Thus the curvature at zero magnetisation changes 
from positive (a minimum) for $H>H_c$ to negative for $H<H_c$.
It is also easy to show that the  solution $m_s$ is always a minimum.
Evaluating the energies at the three solutions 
$m=0, m=m_s, m=1$ yields a behavior typified by Fig.\ 1a.  For 
$H>H_c$, the ground state has $m=0$ and the system is a paramagnet,
for $1\le H \le H_c$, the ground state has $m=m_s$ and is magnetised,
while for $H<1$, the magnetisation saturates.  This behavior is 
summarised in the phase diagram of Fig.\ 1b.  The critical exponent
$\beta =1/y$ on this upper curve in this figure 
and is clearly {\it non-universal in the ground state}.

However when the transition is continuous, 
Aharony\cite{aharony} has demonstrated that at finite temperature 
$\beta = 1/2$ based on the mean-field equation for the random-field
Ising model,
\begin{equation}
m = \int_{-\infty}^{\infty} dh P(h) {\rm tanh}(m/T + h/T). 
\end{equation}
We now reconcile the ground state result (9) found above with the finite
temperature behavior found from equation (11).   Assuming that $P(h)$ is
symmetric, equation (11) can be reduced to,
\begin{equation}
m = 2 {\rm tanh}(2m/T) \int_0^{\infty} { P(h) dh \over 1 + {{\rm cosh}(2h/T)\over {\rm cosh}(2m/T)})}.
\end{equation}
From this expression, it is seen that there are two regimes, $m/T >> 1$ and $m/T<<1$.
At zero temperature only the first regime holds, while at any finite temperature
the second regime is applicable very close to the critical point. 

 When $m/T>>1$,
${\rm tanh}(m/T) \rightarrow 1$  and
 ${\rm cosh}(2h/T)/{\rm cosh}(2m/T) \rightarrow {\rm exp}(2 (h-m)/T)$,
which yields,
\begin{equation}
m = 2 \int_0^{\infty} {P(h) dh \over 1 + {\rm exp}(2(h-m)/T)} \ \ \ m/T \rightarrow \infty.
\end{equation}
Now note that this expression looks like a Sommerfeld integral for the free
fermi gas, with the Fermi energy given by, $\epsilon_f = m$.  The leading 
term at low temperatures is then the integral of $P(h)$ up to the Fermi energy, and 
hence is equivalent to the ground state result given in Eq.\ (6).

However at any finite temperature, there is a regime in which
the magnetisation is small compared to the temperature, $m/T <<1$.
In that case, 
${\rm cosh}(2m/T) \rightarrow 1$, and equation (11) reduces to the mean field
theory for the thermal transition, but with a renormalised coefficient which
depends on the field distribution, ie.,
\begin{equation}
m = 2 I(H,T) {\rm tanh(2m/T) } \ \ \ m/T \rightarrow 0
\end{equation}
where,
\begin{equation}
I(H,T) =  \int_0^{\infty} { P(h) dh \over 1 + {\rm cosh}(2h/T)}.
\end{equation}
Note that there is a factor of two difference in the
argument of the tanh as compared to the thermal mean field
theory.  However the critical temperature and critical exponent
are the same.
For any finite temperature, provided $m<<T$,
an expansion to third order in $m$ of Eq.\ (14) shows that 
the magnetisation approaches zero with exponent $\beta = 1/2$. 
 Moreover, Eq.\ (14) with (15) shows that the 
critical field and temperature are related to each other through the
relation,
\begin{equation}
T = 4 I(H_c(T),T),
\end{equation}
provided the magnetisation is continuous at the transition.  
  For the probability distribution
 given in Eq.\ (8), this reduces to,
\begin{equation}
{H_c(T)\over H_c(0)} =  {\rm tanh}({H_c(T)\over T})-
 ({T \over 2H_c(T)})^y 
\int_0^{2H_c(T) \over T} { x^{y}dx \over 1 + {\rm cosh}(x)}
\end{equation}
The magnetisation as a function of field is given in Fig.\ 2a for
$y=1/2$.  From this figure it is seen that the critical exponent
in the ground state is different than that at finite temperatures,
and clearly illustrates the fact that temperature is a relevant
variable.  The temperature-field phase diagram is presented in
Fig.\ 2b for the two cases $y=1/2$ and $y=2$.  There is 
a sharp shift in the phase boundary with temperature
for cases where $y$ is small (rapidly decaying field distributions 
near the origin), which is strong indicator that temperature is
relevant. 

We have demonstrated the failure of universality in the
ground state of the mean field theory of the 
random field Ising model.  In addition
the concept of a zero temperature fixed point is invalid.  
The fact that the Gaussian distribution of random fields
does have exponent $\beta = 1/2$ in the ground state is
atypical and should not be expected unless the disorder
distribution is quadratic near the origin.  Finite temperature
introduces thermal fluctuations which are also Gaussian which
is the reason that the Gaussian distribution 
of disorder is special and atypical. 

At first blush, our results raise  
serious questions about scaling theories of disordered systems based on
a zero temperature fixed point, and about the applicability
of numerical studies in the ground state to finite temperature
properties.  However this may not be the correct conclusion.
Instead the conventional mean-field
 theory described here may be pathological
and not typical of the behavior in finite dimensions. That
itself would be a rather surprising result, which
for example could be due to a renormalisation
of the disorder distribution to a Gaussian under
rescaling in finite dimensions.  These issues can
only be resolved by careful studies of universality
to disorder in finite dimensions, which is
a difficult task except at zero temperature
where exact numerical calculations are possible.

This work has been supported by
the DOE under contract DE-FG02-90ER45418.


\begin{figure}[f]
\centerline{\epsfig{file=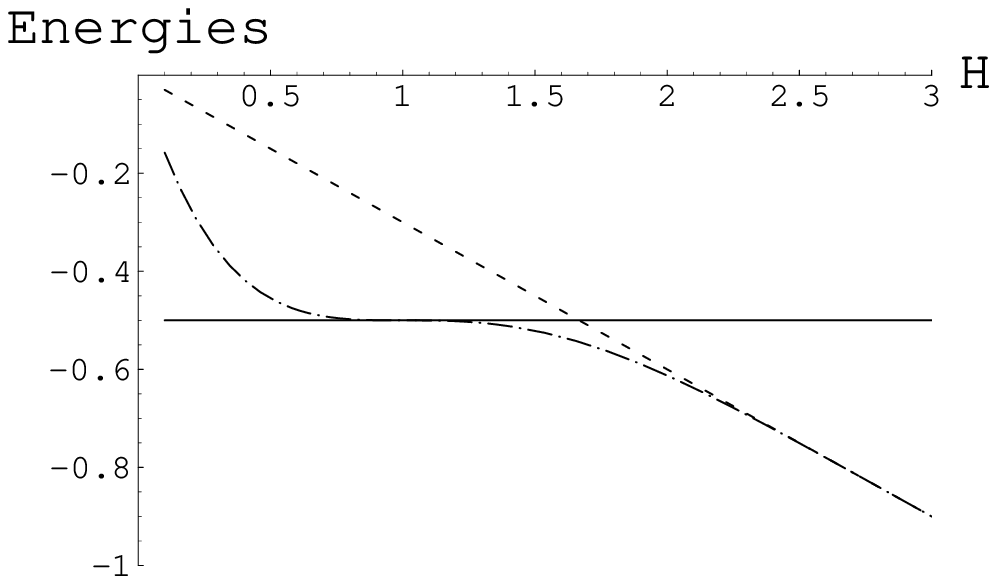,width=7cm}}
\centerline{\epsfig{file=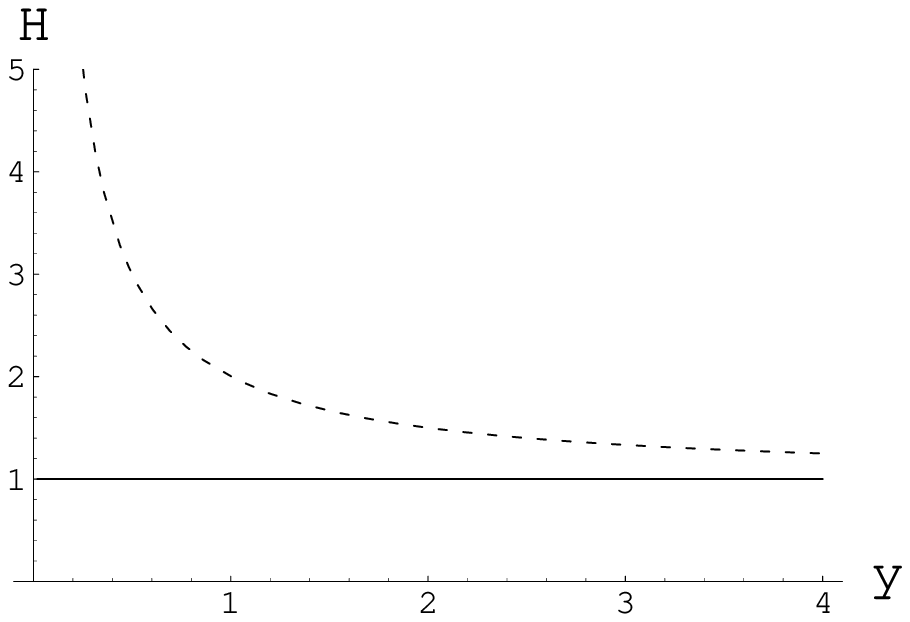,width=7cm}}
\caption{(a) The ground-state energy as a function of the width of the magnetic field distribution ($H$), for the case $y=1/2$.  The flat curve is for 
$m=1$, the linearly decreasing curve is for $m=0$ while the third curve is 
for the solution $m_s$ given in Eq.\ (9) of the text. 
 (b) The upper curve is the dependence of the critical field on the exponent in the
 field distribution (i.e. $H=(y+1)/y$).  Above this 
 line the magnetisation is zero. 
 The lower line is $H=1$, below which the  magnetisation is saturated (i.e. $m=1$ for $H<1$).  Between these two lines the magnetisation obeys Eq.\ (9) of the text, with the magnetisation going to zero with exponent $1/y$ at the upper curve.}
\label{fig1}
\end{figure}

\begin{figure}[f]
\centerline{\epsfig{file=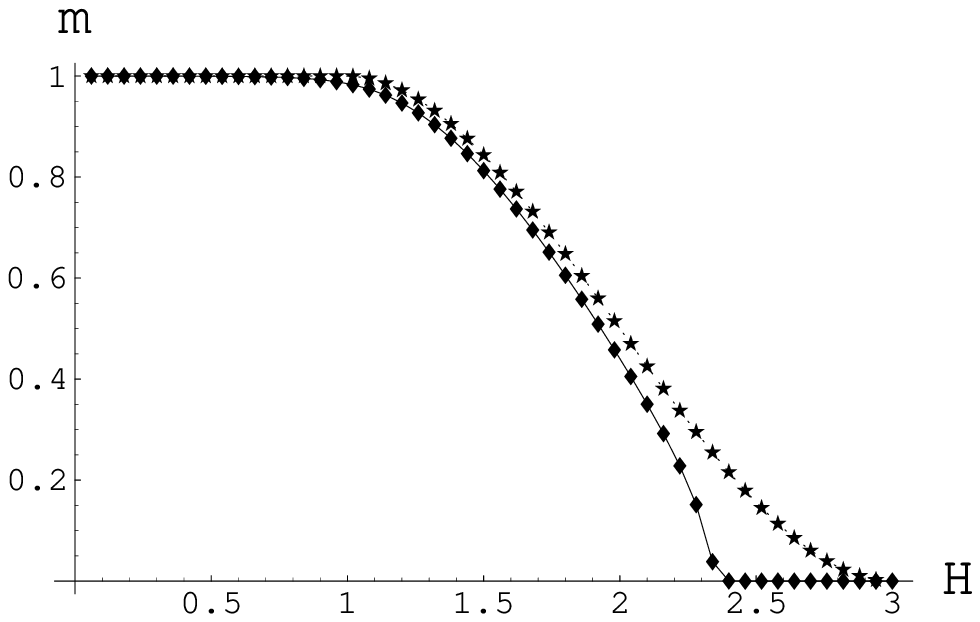,width=7cm}}
\centerline{\epsfig{file=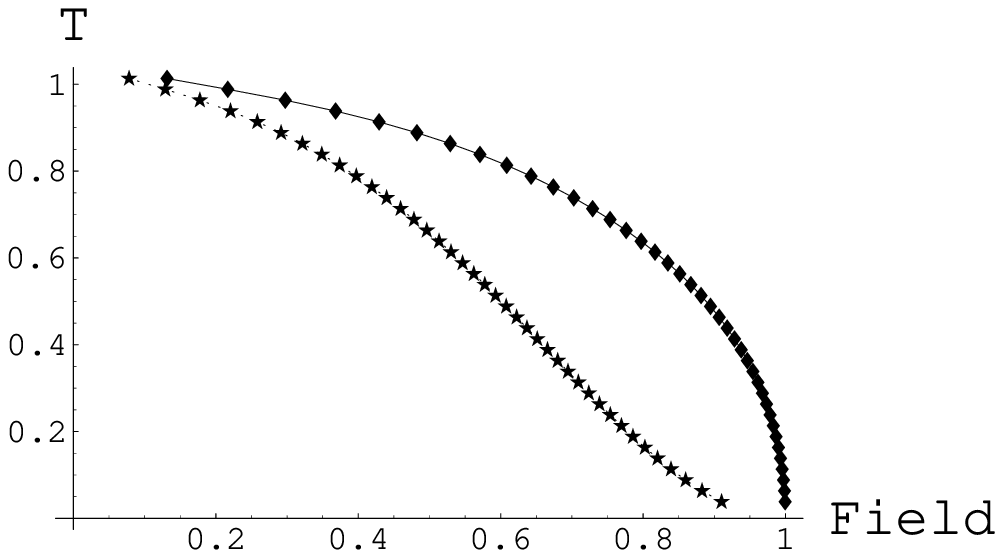,width=7cm}}
\caption{(a) The magnetisation as a function
of the width of the random field ($H$) for $y=1/2$, for two temperatures.
The upper curve is for zero temperature, while the lower curve
is for $T=0.2$.
 (b) The $H-T$ phase diagram ($T$ vs. $H_c(T)/H_c(0)$) for $y=2$ (upper curve) and $y=1/2$ (lower curve) found from solving Eq.\ (17) of the text.}
\label{fig2}
\end{figure}
\end{multicols}
\end{document}